\documentclass[times, tighten, twocolumn]{aastex631} 

\usepackage[T1]{fontenc}
\usepackage{array}
\usepackage{amsmath}
\usepackage{multirow}

\hypersetup{
linkcolor=magenta
}

\graphicspath{{./}{figure/}}

\newcolumntype{x}{>{\color{blue}}c}

\newcommand{\fluxunit}{\rm\, ph\,s^{-1}\, cm^{-2}}
\newcommand{\emin}{E_{\rm min}\,=\, 100\, \rm MeV}

\newcommand{\emax}{E_{\rm max}\,=\, 300\, \rm GeV}
\newcommand{\LnuI}{L_{\rm 1.4\,GHz}}

\newcommand{\xinth}{\xi_{\rm nth}}

\newcommand{\Lgam}{L_{\rm 0.1-300\,GeV}}
\newcommand{\Lgamp}{L_{\rm 0.1-300\,GeV}^\prime}

\newcommand{\Mbh}{M_{\bullet}}
\newcommand{\Msun}{M_{\odot}}

\newcommand{\Lbol}{L_{\rm Bol}}

\newcommand{\LIR}{L_{\rm 8-1000 \,\mu m}}
\newcommand{\Lsoftx}{L_{\rm 0.3-10\,keV}}
\newcommand{\Lx}{L_{\rm 14-195\,keV}}
\newcommand{\Rg}{R_{\rm g}}
\newcommand{\Rrx}{\mathcal{R}_{\rm rX}}
\newcommand{\RrB}{\mathcal{R}_{\rm rB}}
\newcommand{\ergs}{\rm\,erg\,s^{-1}}
\newcommand{\Rcom}{R} 
\newcommand{\taugg}{\tau_{\gamma\gamma}}
\newcommand{\sigmagg}{\sigma_{\gamma\gamma}}
\newcommand{\sigmaT}{\sigma_{\rm T}}

\newcommand{\zetanu}{\zeta{(\nu)}}
\newcommand{\Lnup}{L_\nu^\prime}
\newcommand{\nep}{n_{\rm e}}
\newcommand{\gammax}{\gamma_{\rm max}}
\newcommand{\gammin}{\gamma_{\rm min}}
\newcommand{\game}{\gamma_{\rm e}}

\newcommand{\me}{m_{\rm e}}

\newcommand{\MVal}{1.66\times10^8}
\newcommand{\LxVal}{2.4\times10^{43}\,\rm erg\, s^{-1}}
\newcommand{\LbolVal}{2.3\times 10^{44}\ergs}
\newcommand{\NHVal}{2.0\times 10^{24}}

\newcommand{\fcVal}{5.1~(\pm 1.5)\times10^{-9}\fluxunit}
\newcommand{\RgamX}{2.4\%}
\newcommand{\TSPL}{42.81} 
\newcommand{\sigPL}{6.22}
\newcommand{\GamPL}{2.61~(\pm 0.24)}
\newcommand{\LgamVal}{5.9\,(\pm 1.7)\times10^{41}}

\newcommand{\RrxVal}{1.3\times 10^{-5}}

\newcommand{\xinthVal}{14\%}
\newcommand{\gamjetVal}{2.0\,(\pm 0.1)\times10^3}
\newcommand{\RjetVal}{2.38\,(\pm0.12)\,\rm pc}
\newcommand{\njetVal}{0.88\,(\pm 0.12)\,{\rm cm^{-3}}}
\newcommand{\BjetVal}{1.53\,(\pm0.12)\times10^{-3}\,\rm G}
\newcommand{\gamcorVal}{2.2\,(\pm 2.1)\times10^5}
\newcommand{\ncorVal}{1.1\,(\pm0.1)\times10^7\,{\rm cm^{-3}}}
\newcommand{\LgampCorVal}{3.4\,\times10^{42}}
\newcommand{\LgampJetVal}{2.2\,\times10^{41}}
\newcommand{\ratioI}{0.07}  
\newcommand{\ratioII}{0.19} 
\newcommand{\ratioIII}{4.3} 

\newcommand{\LIRVal}{6.0\times 10^{10}\,L_\odot}
\newcommand{\LgamSFG}{L_{\gamma,\rm\,SFG}={2.9\,(\pm 0.6)\times 10^{40}\ergs}}

\def\ihep{State Key Laboratory of Particle Astrophysics, Institute of High Energy Physics,
Chinese Academy of Sciences, 19B Yuquan Road, Beijing 100049, China}

\def\AstroUCAS{School of Astronomy and Space Sciences, University of Chinese Academy of Sciences, 
19A Yuquan Road, Beijing 100049, China}

\def\naoc{National Astronomical Observatory of China, 20A Datun Road, Beijing 100020, China}

\shorttitle{Fermi detection of $\gamma$-rays from NGC 3281}
\shortauthors{Liu et al.}

\begin{document}

\title{Fermi detection of $\gamma$-rays from the radio-quiet Seyfert galaxy NGC 3281}

\author[0000-0003-3086-7804]{Jun-Rong Liu}
\affil{\ihep}

\author[0000-0001-7584-6236]{Hua Feng}
\affil{\ihep}
\email{hfeng@ihep.ac.cn}

\author[0000-0001-9449-9268]{Jian-Min Wang}
\affil{\ihep}
\affil{\AstroUCAS}
\affil{\naoc}

\begin{abstract}
We report the detection of significant $\gamma$-ray emission with $\it Fermi$-LAT from the radio-quiet Seyfert 2 galaxy NGC 3281, with a luminosity of $5.9\,(\pm 1.7)\times10^{41}\rm\,erg\,s^{-1}$ at a significance of $6.22\,\sigma$ (TS = $42.81$). The power-law photon index is $2.61~(\pm 0.24)$, indicative of a soft spectrum. 
The star formation activity in NGC 3281 is insufficient to explain its $\gamma$-ray luminosity based on the empirical relation between the infrared and $\gamma$-ray luminosities observed in other sources.
The multiwavelength spectrum can be explained as due to inverse Compton scattering by relativistic electrons in the corona or jet of seed photons from the corona, disk and torus. 
The source is Compton-thick and attenuation of GeV photons due to pair production in the corona is nonnegligible (with an optical depth of about 10). 
The intrinsic $\gamma$-ray luminosity is inferred to be $3.4\,\times10^{42}$ and $2.2\,\times10^{41}\rm\,erg\,s^{-1}$ for the corona and jet model, respectively.
The observed $\gamma$-ray and radio luminosities is roughly consistent with the known correlation between the two quantities, among the lowest luminosity regime.
The jet origin is valid only if the radio emission is dominated by the jet. 
\end{abstract}



\section{Introduction}

Radio-loud active galactic nuclei (AGNs) can display strong $\gamma$-ray emission, due to inverse Compton scattering by electrons in the relativistic jet \citep[e.g., see review in][]{Blandford2019}.
GeV emission is also seen in some radio-quiet AGNs.
For example, AGNs with a strong star formation rate (SFR) are often significantly detected in $\gamma$-rays \citep{Ackermann2012_SF, Ajello2020}.
\citet{Liu2025} found evidence for $\gamma$-ray emission in the corona of AGNs from a carefully selected sub-sample in the Swift BAT AGN Spectroscopic Survey (BASS).
These are nearby, radio-quiet, and ultra hard X-ray bright AGNs, whose jets are not powerful enough to explain the observed GeV emission.
There are also signatures of $\gamma$-ray emission below the GeV band arguably originated in the AGN corona from NGC 1068 \citep{Ajello2023} and NGC 4945 \citep{Murase2024}.

The AGN corona is expected to be a factory of $\gamma$-ray emission \citep{Inoue2019, Inoue2021} in light of its non-thermal activity found in radio observations \citep{Inoue2018}.
\cite{Bhattacharyya2003, Bhattacharyya2006} have found that the proton-proton collisions in the inner region of an accretion disk would produce electrons, positrons, and $\gamma$-rays.
An ultra high energy bump above TeV is expected due to injection of non-thermal particles into a magnetized corona \citep{Romero2010}.

$\gamma$-rays from AGN coronae have been searched in the past.
Observations with the Energetic Gamma Ray Experiment Telescope (EGRET) gave a $2\sigma$ upper limit on the $\gamma$-ray flux from 22 Seyfert galaxies selected by X-ray \citep{Lin1993}. 
\cite{Cillis2004} utilized a stacking technique on a sample of Seyfert galaxies away from the Galactic plane and found no significant result ($<2\sigma$) with EGRET data.
\cite{Teng2011} analyzed the $\it Fermi$ Large Area Telescope (LAT) data on 491 Seyfert galaxies detected by Swift BAT and found that only two nearby objects, NGC 1068 and NGC 4945, were detected with $\gamma$-ray emission, consistent with starburst activity.
For those undetected objects, \cite{Teng2011} gave a flux upper limit $\sim 2 \times 10^{-11} \fluxunit$ above 1 GeV.
\cite{Ackermann2012_Seyfert} searched $\it Fermi$-LAT data of 120 hard X-ray-selected Seyfert galaxies again and found no significant $\gamma$-ray emission in the GeV band.
However, only 2.1 and 3 years of the $\it Fermi$-LAT data were analyzed in \cite{Teng2011} and \cite{Ackermann2012_Seyfert}, respectively.
Recently, sub-GeV $\gamma$-rays from NGC 4945 were reported by \cite{Murase2024} and explained as due to the magnetically powered corona model, while its GeV emission was explained by star formation activity.
In a similar case of NGC 1068, a two-component model, with hadronic emission in the vicinity of the central supermassive black hole (SMBH) and starburst activity, is proposed to explain the low-energy (below $\sim 500$~MeV) and high-energy (above $\sim 500$~MeV) $\gamma$-ray spectral energy distribution (SED), respectively \citep{Ajello2023}.
Another radio-quiet galaxy, NGC 4151, may have shown $\gamma$-ray emission, which can be explained as due to jet, coronal activities \citep{Inoue2023}, or the ultra fast outflows \citep{Ajello2021}, despite the fact that for this 
object \citep[NGC 4151, see][]{Peretti2023} the origin of its $\gamma$-ray emission is uncertain since there is a blazar within the error radius of the $\gamma$-ray position \citep{Ballet2023,Murase2024}.

In this work, we studied the $\gamma$-ray properties of another radio-quiet Seyfert galaxy, NGC 3281, using $\it Fermi$-LAT data accumulated in around 16 years.
The target was selected from a $\gamma$-ray survey of the BASS sample \citep{Liu2025}.
NGC 3281 is a spiral Seyfert 2 galaxy (R.A.\ $=157\fdg9670$, Decl.\ $=-34\fdg8537$) with a redshift based distance of $D=48.1$ Mpc \citep{Koss2022}.
It is detected by Swift-BAT with a hard X-ray luminosity of $\Lx = \LxVal$  \citep{Oh2018}
and by Swift-XRT with a soft X-ray luminosity of $\Lsoftx=9.7\times10^{41}\ergs$ \citep[][]{Evans2020}.
Its nuclear region is Compton-thick with a column density of $N_{\rm H}=\NHVal\,\rm cm^{-2}$ \citep{Zhao2021}.
The mass of SMBH is $\Mbh=\MVal\,\Msun$ derived from velocity dispersion \citep{Koss2022b}.
The bolometric luminosity is derived as $\Lbol=\LbolVal$, assuming a 14-150 keV bolometric correction of 8 \citep{Ricci2017}, consistent with that derived based on [O~{\sc iii}] \citep{Dall2023}.

This paper is organized as follows.
We describe the $\it Fermi$-LAT data analysis and present the results in \S\,\ref{sect:data analysis}.
We explain the $\gamma$-ray emission in \S\,\ref{sect:model}, with the corona model (\S\,\ref{sect:corona}) and the jet model (\S\,\ref{sect:jet}).
We discuss the results in \S\,\ref{sect:discussion}.

\section{Data analysis and results}
\label{sect:data analysis}

The LAT on the ${\it Fermi\, Gamma}$-${\it ray\, Space\, Telescope}$ scans the whole sky approximately every 3 hours to take full advantage of its large field of view of 2.4~sr \citep{Atwood2009}.
We extracted and analyzed the $\it Fermi$-LAT all-sky weekly data from modified Julian date (MJD) 54682 to MJD 60593 (from August 4, 2008 to October 10, 2024) accumulated in around 16 years using the 
{\tt Fermipy} (v1.1.6) packages \citep{Wood2017}.
First, we selected source class events with the tool $\tt{gtselect} $ by setting $\tt evclass=128$ and $\tt evtype= 3$.
Photons within 15\arcdeg\ are selected to account for the large point spread function at lower energies (below 1~GeV).
Second, the standard $\tt gtmktime$ filter recommended by the LAT team was used to select good time intervals and valid data by setting $\tt(DATA\_QUAL>0)\&\&(LAT\_CONFIG==1)$. 
Photons are divided into 30 logarithmic energy bins between $\emin$ and $\emax$ by setting $\tt enumbins=30$.

The Galactic background file $\tt gll\_iem\_v07$ and isotropic diffuse background file $\tt iso\_P8R3\_SOURCE\_V3\_\tt v1$ were considered in the fit, with both the normalization and index as free parameters.
Point sources from 15$^\circ$ to 20$^\circ$ around NGC 3281 in the $\tt gll\_psc\_v32$ file \citep[the 4FGL-DR4 catalog;][]{Ballet2023} are included, with all model parameters fixed in the fit.
The normalizations for point sources from 5 to 15$^\circ$ with TS~$>500$ are set free.
For point sources within $5^\circ$, their flux and spectral index are set to be free parameters. 
The energy dispersion correction for each energy bin is implemented by setting $\tt edisp\_bins=-1$.
We performed a likelihood fit of the $\gamma$-ray spectrum with a power-law model assuming Poisson statistics\footnote{See description of the likelihood at \url{https://fermi.gsfc.nasa.gov/ssc/data/analysis/documentation/Cicerone/Cicerone_Likelihood/} and analysis threads at \url{https://fermi.gsfc.nasa.gov/ssc/data/analysis/scitools/}.}.
The $\tt MINUIT$ optimizer is employed with a tolerance of $10^{-4}$ \citep{James1975}.

\begin{deluxetable}{cccc}
\footnotesize
\tablecaption{$\gamma$-ray flux and TS in each energy bin.}
\label{tab:sed_data}
\tablehead{
\colhead{$E_{\rm min}$} & \colhead{$E_{\rm max}$} & \colhead{flux} & \colhead{TS} \\
\colhead{(GeV)} & \colhead{(GeV)} & \colhead{($10^{-13}\rm\, erg\,s^{-1}\,cm^{-2}$)}  & \colhead{}}
\startdata
$0.10$&$0.27$&$8.9\pm 2.5$&13.8 \\
$0.27$&$0.74$&$7.4\pm 1.4$&29.5 \\
$0.74$&$2.0$&$2.89\pm 0.82$&8.34 \\
$2.0$&$5.5$&$<1.98$&1.63\\
$5.5$&$15$&$1.00\pm0.62$&5.94 \\
$15$&$41$&$<0.84$&0 \\
$41$&$110$&$<2.28$&0 \\
$110$&$300$&$<9.09$&0 \\
\enddata
\end{deluxetable}

\begin{figure}[hbt]
\centering
\includegraphics[width=0.8\linewidth]{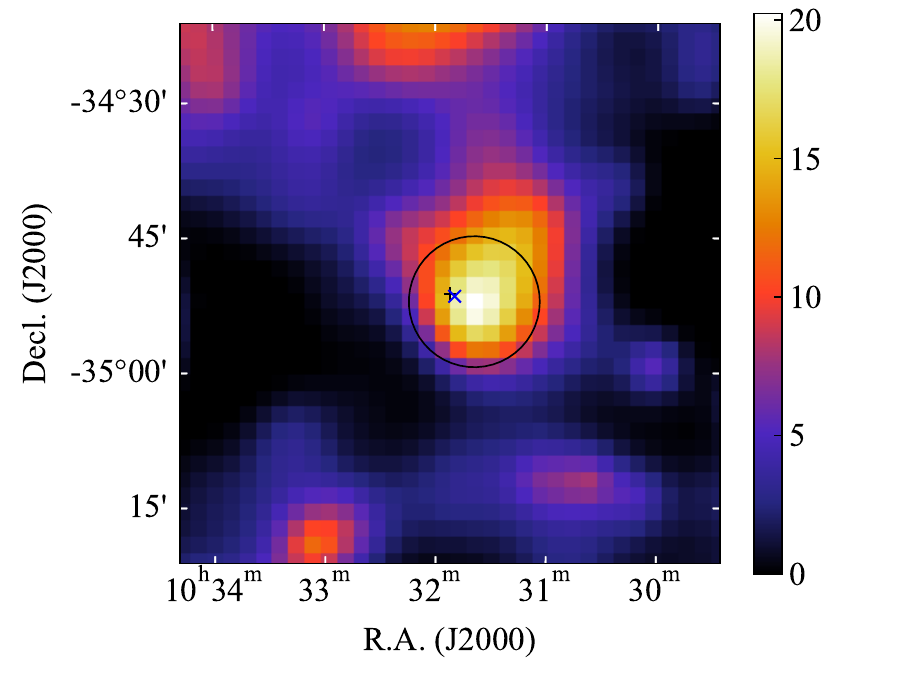}\\
\includegraphics[width=0.8\linewidth]{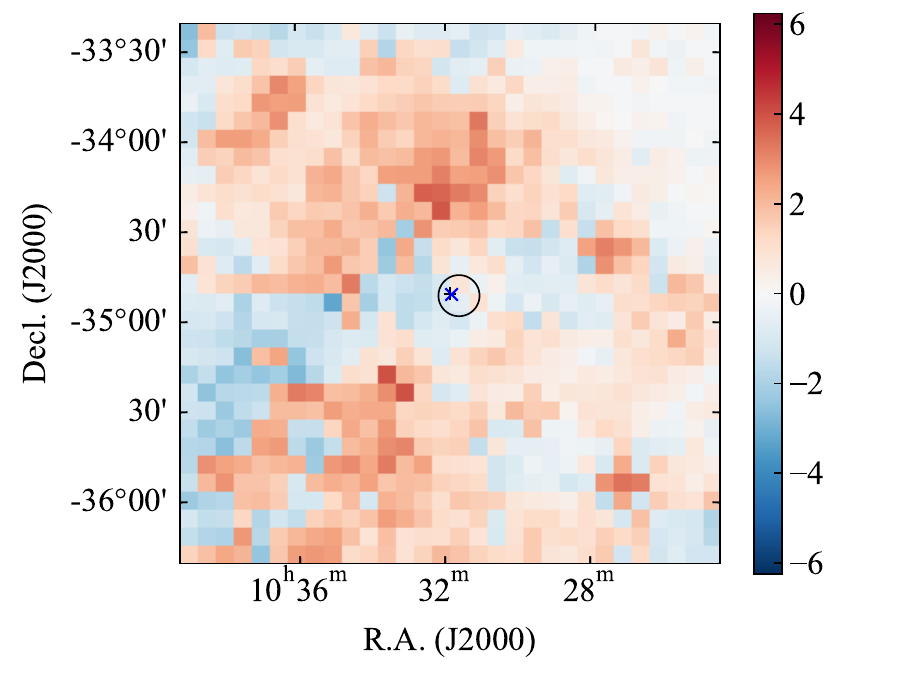}
\caption{
TS (top) and PS (bottom) maps around NGC 3281. 
The black plus and blue cross indicate the optical and X-ray position of NGC 3281, respectively.
The black circle marks the 95\% error circle of the $\gamma$-ray localization.
The color bars indicate the range of the TS/PS values.
The TS map has a pixel size of 0\fdg03 and shows a point-like source consistent with the optical/X-ray position.
We note that the maximum TS is different from that reported in the text (42.73) as the power-law photon index is fixed at 2 in generating the map.
The PS map has a pixel size of 0\fdg1 and shows no value above 
6.24 ($5\sigma$).}
\label{fig:TSmap}
\end{figure}

\begin{figure}
\centering
\includegraphics[width=\linewidth]{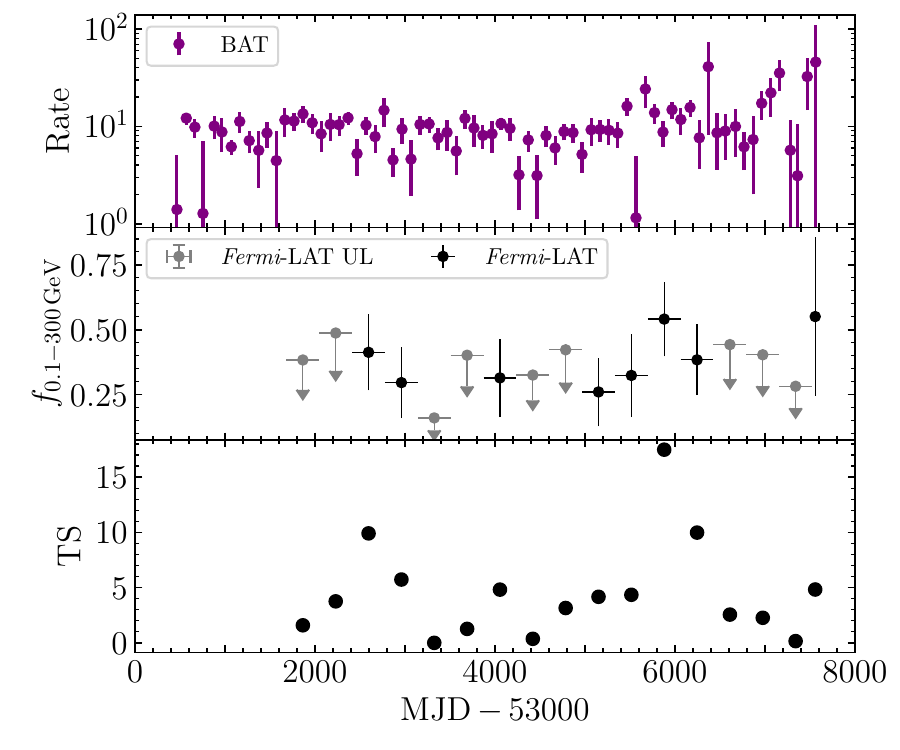}
\caption{X-ray and $\gamma$-ray light curves of NGC 3281.
The BAT light curve in the unit of $10^{-4}\,\rm counts\,s^{-1}$ is binned at a step of 100~d.
The LAT light curve in the unit of $10^{-11}\,\rm erg\,s^{-1}\,cm^{-2}$ is binned at a step of a year.
Gray arrows represent the 95\% upper limit if $\rm TS<4$.
}
\label{fig:lc_multi}
\end{figure}

The $\tt fit$ tool is used to fit the data and obtain the TS value.
The TS and PS maps are produced using the $\tt tsmap$ and $\tt psmap$ tools, respectively, to check the source detection and background residuals, as shown in Figure \ref{fig:TSmap}. 
The $\gamma$-ray position with error radius is derived using the $\tt localize$ tool.
We used the $\tt lightcurve$ tool to obtain the $\gamma$-ray light curve at a bin size of a year (Figure~\ref{fig:lc_multi}).
Also, we divided the energy band of 100 MeV-300 GeV into 8 logarithmic bins to derive a SED (Table~\ref{tab:sed_data}) using the $\tt sed$ tool.

\begin{figure}
\centering
\includegraphics[width=\linewidth]{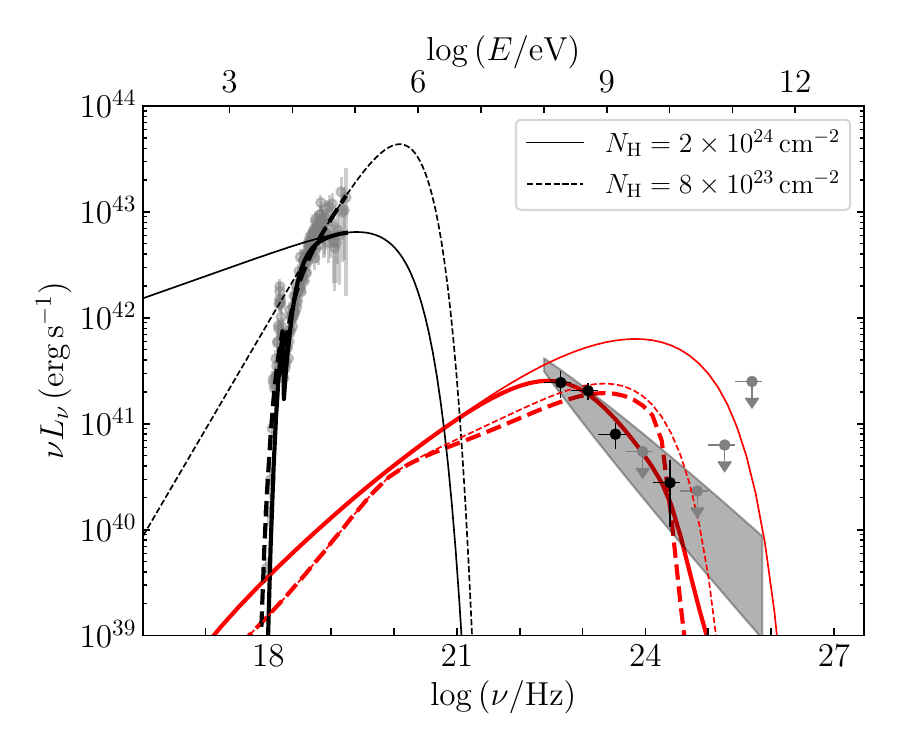}
\caption{The X-ray  and $\gamma$-ray spectra and best-fit models assuming two different $N_{\rm H}$ in the context of the corona model.
The thin and thick lines represent the intrinsic and attenuated models, respectively.}
\label{fig:sed_corona}
\end{figure}

\begin{figure*}
\centering
\includegraphics[width=\linewidth]{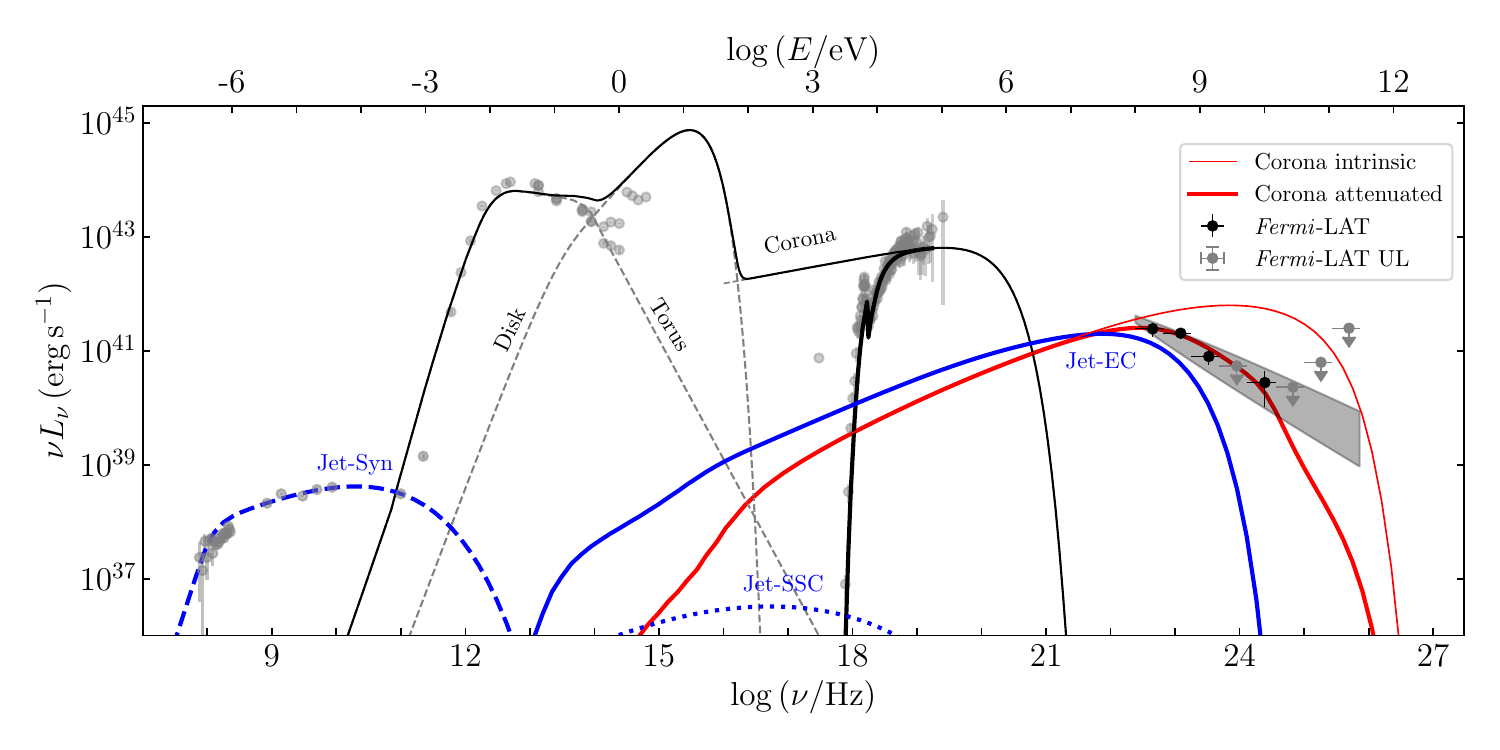}
\caption{
Multiwavelength SED of NGC 3281 with best-fit model components.
The red curves represent the intrinsic (thin) and observed (thick) model flux due to inverse Compton scattering in the corona.
The blue curves represent the external inverse Compton (solid), synchrotron (dashed), and synchrotron self-Compton (dotted) processes in the jet.
The black curves represent the SEDs of seed photons. 
The radio \citep{Ulvestad1989,Condon1998,Schmitt2001,Mauch2003,Gordon2021,Ricci2023,Kawamuro2023,Ross2024}, infrared \citep{Sanders2003,Melendez2014,Shimizu2016,Marocco2021}, optical-UV \citep{Skrutskie2006,Koss2011}, and X-ray data \citep{Cardamone2007,Ricci2017,Oh2018,Balokovic2020} are collected from the archive.
}
\label{fig:sed}
\end{figure*}

The $\gamma$-ray localization has $\text{R.A.} = 157\fdg91$ and $\text{decl.} = -34\fdg86$ with a 95\% error radius of 0\fdg12, consistent with the optical and X-ray positions of NGC 3281 (see Figure \ref{fig:TSmap}).
Here the systematic correction is considered \citep[see Equation 1 in][]{Abdollahi2020}.
We searched for blazars in the $\gamma$-ray error radius against two blazar catalogs, the Roma-BZCAT catalog \citep{Massaro2015} and the WISE Blazar-like Radio-Loud Sources (WIBRaLS) catalog \citep{Abrusco2019}, and found no matches. 

NGC 3281 is detected with a TS value of  $\TSPL\,(\sigPL\,\sigma)$. 
The best-fit power-law photon index is $\GamPL$, and the integrated count flux from 100 MeV to 300 GeV is $\fcVal$, corresponding to a $\gamma$-ray luminosity $\Lgam=\LgamVal\ergs$ at a distance of 48.1 Mpc.
No significant variability is seen on the timescale of one year, with $\text{TS}_{\rm var}=15.53$ ($0.70\,\sigma$) tested against a constant flux.
The source may have displayed a flare in both the $\gamma$-ray and X-ray bands around MJD 59000, but the significance is low.

\section{Modeling}
\label{sect:model}

We try to interpret the $\gamma$-ray emission in NGC 3281 as a result of inverse Compton scattering by high energy electrons in the corona and jet, respectively.

\subsection{Corona model}
\label{sect:corona}

In an AGN corona, electrons can be accelerated approaching the speed of light by shocks \citep{Drury1983,Blandford1987, Inoue2019} or magnetic reconnection \citep{Sironi2014, Chael2017}.
The energy spectrum of the non-thermal electrons is given by 
\begin{equation}
\frac{{\rm d}N}{{\rm d}\game} = n_0 \game^{-p}e^{-\game / \gammax} \; ,
\label{eq:electron}
\end{equation}
where $\game$ is the Lorentz factor,
$\gammax$ is the cutoff Lorentz factor,
$p$ is the spectral index assumed to be 2,
and $n_0$ is the normalization.
The total number density  $\nep$ can be obtained by integrating ${\rm d}N/{\rm d}\game$ from $\gammin = 1$ to infinity.

Emission from the corona, disk, and torus provides a seed photon field for inverse Compton scattering ($\gamma$-ray production) as well as pair production ($\gamma$-ray attenuation).
The corona emission dominates in the X-ray band, typically in the energy range of $\sim$0.1-100 keV, and can be described by a power-law distribution with an exponential cutoff, $L_{\nu,\rm cor}=L_0 (E / E_{\rm c})^{\alpha}e^{-E/E_{\rm c}}$, where $\alpha$ is the spectral index, $E_{\rm c}$ is the cutoff energy, and $L_0$ is the normalization. 
The corresponding number density of photons is $n_{\rm ph}(\nu)=L_{\nu,\rm cor}/(4\pi \Rcom^2 h\nu c)$, where $h$ is the Planck constant,
$c$ is the speed of light,
and $\Rcom$ is the corona radius. 
We adopt $\Rcom = 10\,\Rg$, where $\Rg=G\Mbh/c^2$ is the gravitational radius and $G$ is the gravitational constant.
For the accretion disk, the standard disk \citep{Shakura1973} model with radiation peaked in optical and UV is assumed. 
A radius of $\Rcom = 10^{3}\,\Rg$ is assumed to calculate the number density of photons.
For the torus, the template spectrum in \cite{Sanders1989} is used. The emission peaks in infrared and is assumed to be confined within a radius of $\Rcom = 10^6\,\Rg$.
The SEDs of the three components are displayed in Figure \ref{fig:sed}. 
Their normalizations are found by fitting with the data.

Pair production attenuates $\gamma$-rays originated in the corona \citep{Inoue2019}. 
Therefore, the model flux can be calculated as a result of inverse Compton scattering by high energy electrons of seed photons  (see Appendix \ref{sect:IC}), subject to attenuation caused by pair production (see Appendix \ref{append:pair}), and then compared with data using the Markov Chain Monte Carlo (MCMC) fit.

For the X-ray fit, we adopted the NuSTAR spectrum previously reported in \citet{Zhao2021}. 
As the NuSTAR spectrum cannot well constrain the absorption column density due to limited low energy coverage, we fixed $N_{\rm H}$ at $8 \times 10^{23}$ and $2 \times 10^{24}\,\rm cm^{-2}$, respectively, the bounds of $N_{\rm H}$ found by \citet{Zhao2021}.
The cutoff energy $E_{\rm c}$ is fixed at 500~keV \citep{Zhao2021}. 
The X-ray and $\gamma$-ray SED and best-fit models are shown in Figure~\ref{fig:sed_corona}.
As one can see, $N_{\rm H}$ at the upper bound provides a better fit in the GeV band than the lower one does.
Thus, we fixed $N_{\rm H} = 2\times10^{24}\,\rm cm^{-2}$ in the subsequent analysis.
The multiwavelength SED and best-fit model is shown in Figure~\ref{fig:sed}.
The X-ray spectral model parameters are derived as $\alpha = -0.79$ and $L_0=9.2\times 10^{22}\ergs\,Hz^{-1}$.
Correspondingly, the differential luminosity at 0.91~keV and integrated luminosity from 2 to 10~keV are $2.9\times 10^{42}\ergs$ and $6.5\times 10^{42}\ergs$, respectively.
The best-fit parameters are listed in Table \ref{tab:sed_para}.

We obtain the intrinsic $\gamma$-rays luminosity $\Lgamp=\LgampCorVal\ergs$ and $\gamma$-ray to X-ray luminosity ratio $\xinth=\Lgamp/\Lx\approx\xinthVal$.
If $\gamma$-rays are a proxy for non-thermal electrons (synchrotron radiation is not considered here) and X-rays for thermal electrons, $\xinth$ then represents the fractional energy of non-thermal electrons.
The observed ratio $\Lgam/\Lx=\RgamX$ is consistent with a previous observational upper limit of 0.1 \citep{Ackermann2012_Seyfert}.

We calculated the optical depth by integrating the pair production cross section in Appendix~\ref{append:pair}.  
The optical depth as well as the transmission fraction as a function of energy are shown in Figure \ref{fig:tau}.
As one can see, the optical depth at 1~GeV is $\sim$10, indicating that the attenuation due to pair production cannot be ignored in our case.
We note that a consistent optical depth can be obtained if one applies Eq.~(1) in \cite{Murase2024} for a simplified estimation.

\begin{deluxetable}{ccc}
\footnotesize
\tablecaption{Best-fit SED parameters of the two models.}
\label{tab:sed_para}
\tablehead{
\colhead{Parameter} & \colhead{Corona} & \colhead{Jet}}
\startdata
$p$ & $2^\ast$ & $2^\ast$ \\
$\gammin$ & $1^\ast$ & $1^\ast$ \\
$\gammax$ & $\gamcorVal$ & $\gamjetVal$ \\
$\Rcom$&$10^\ast R_{\rm g}$&$\RjetVal$ \\
$\nep$ &$\ncorVal$&$\njetVal$ \\
$B$ & \nodata & $\BjetVal$ \\
\enddata
\tablenotetext{\ast}{fixed in the fit.}
\end{deluxetable}

\subsection{Jet model}
\label{sect:jet}

\begin{figure}
\centering
\includegraphics[width=0.9\linewidth]{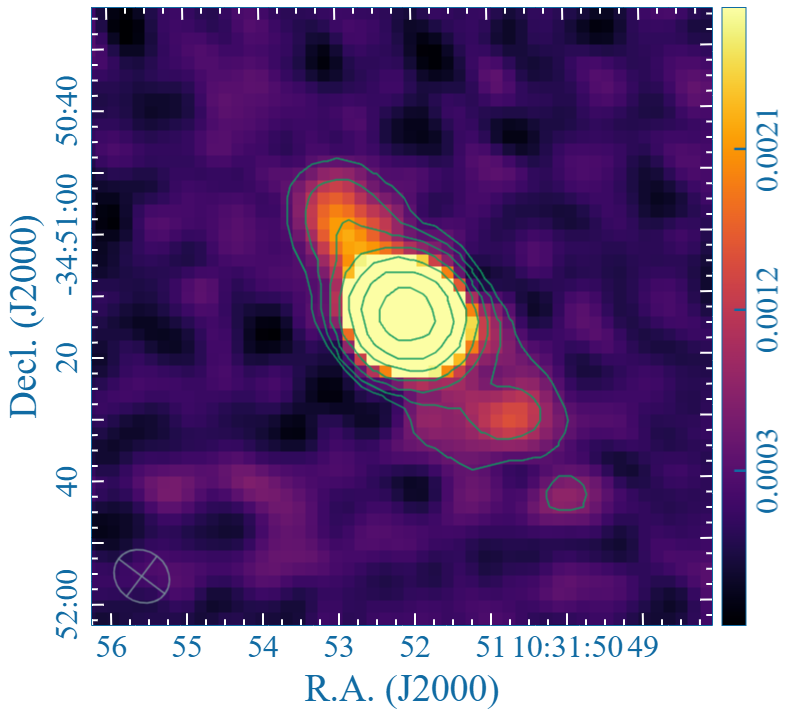}
\caption{ASKAP image at 1.4 GHz. The contours are 3, 6, 12, 24, 48, 96, and 192 $\sigma$ (0.144 mJy/beam), respectively.}
\label{fig:ASKAP}
\end{figure}

NGC 3281 exhibits a radio core with extended jet-like features, as shown in a 1.4~GHz image obtained with the Australian Square Kilometre Array Pathfinder \citep[ASKAP;][]{Johnston2008} in Figure~\ref{fig:ASKAP}.

Given the 5 GHz and $B$ band luminosity densities \citep{Ulvestad1989,Veron2010}, the source has a radio-to-optical luminosity ratio $\RrB \equiv L_{\nu,\rm 5GHz} / L_{\nu,\rm B} = 7.3$ \citep{Ulvestad1989,Veron2010}, classified as a radio-quiet AGN \citep[$\RrB < 10$;][]{Kellermann1994}.
As $\RrB$ is significantly influenced by the subtraction of host galaxy emission, especially for Seyfert galaxies \citep{Ho2001}, 
we also calculated the radio loudness against the hard X-ray band, defined as $\Rrx \equiv L_{\rm 1.4GHz}/\Lx=\RrxVal$ \citep{Teng2011,Ackermann2012_Seyfert}, which also aligns with the classification of a radio-quiet AGN, characterized by $\Rrx < 10^{-4}$ \citep{Ackermann2012_Seyfert} or $\Rrx < 2.0\times10^{-5}$ \citep{Teng2011}.
Here $\LnuI$ is adopted from \citet{Condon1998}.

A simple one-zone homogeneous jet model \citep{Inoue1996} is adopted to calculate the radio and $\gamma$-ray SEDs, assuming the same power-law spectral distribution described in Eq.~(\ref{eq:electron}).
We adopted the same seed photon field outlined in \S\, \ref{sect:corona}.
The jet radius, referring to the distance to the SMBH, is set as a free parameter in the fit. 
The seed photon density at the jet radius is calculated assuming an inverse square law if the component radius is smaller than the jet radius. 
The synchrotron radiation and synchrotron self-Compton scattering  are calculated following \cite{Inoue1996} and the external inverse Compton scattering is derived from Eq.~(\ref{eq:Lnup}).
The pair production process is unimportant and ignored in this case with a large jet radius.
The radio and $\gamma$-ray SEDs are fitted simultaneously.
The fit includes four free parameters, $n_0$, $\gammax$, $R$ (jet radius), and $B$ (jet magnetic field strength).
In principle, these parameters can be determined from the synchrotron self-absorption frequency, peak frequency of synchrotron radiation, radio luminosity, and $\gamma$-ray luminosity. 
The best-fit results are also listed in Table~\ref{tab:sed_para}. 
The intrinsic $\gamma$-ray luminosity in the jet model is $\LgampJetVal\ergs$.

\begin{figure}
\centering
\includegraphics[width=\linewidth]{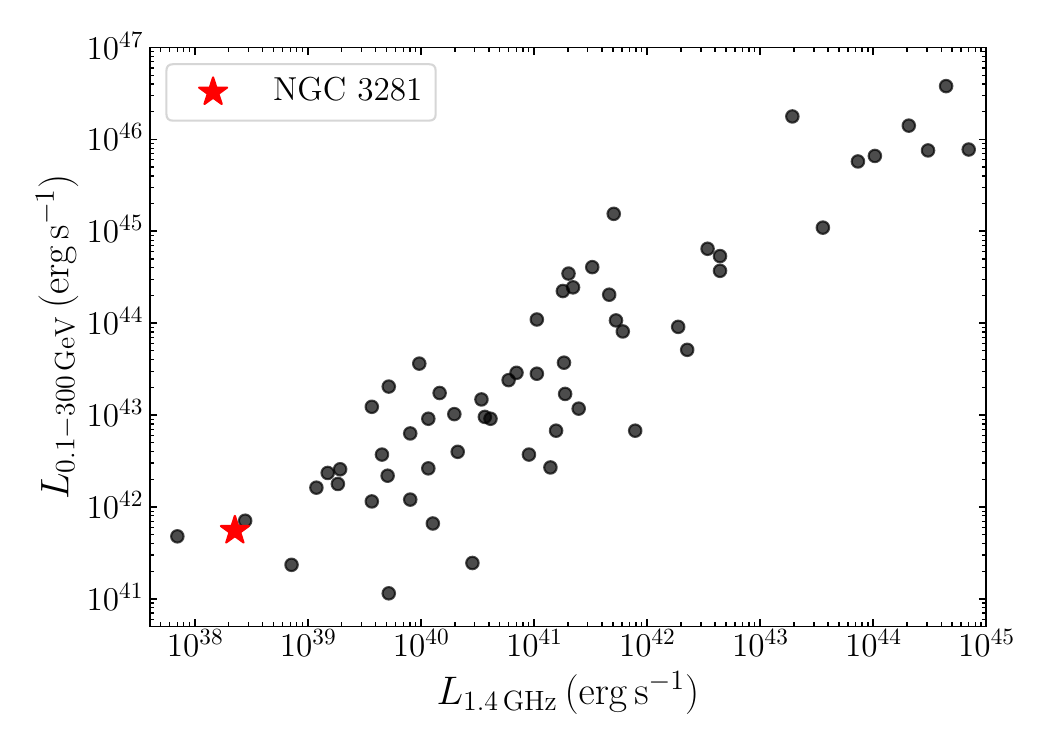}
\caption{Gamma-ray vs.\ radio luminosities for AGNs in \citet{Fukazawa2022} and this source.
}
\label{fig:rgcorr}
\end{figure}

\section{Discussions}
\label{sect:discussion}

Here we report a significant detection of  $\gamma$-ray emission from the radio-quiet AGN in NGC 3281 with $\it Fermi$-LAT, with $\Lgam=\LgamVal\ergs$ at a significance of $\sigPL\,\sigma$.
The $\gamma$-ray emission, as well as the multiwavelength SED, can be explained in the context of inverse Compton scattering by high energy electrons in the corona or in the jet of seed photons from the torus, disk and corona. 
In the corona model, the seed photons mainly come from the corona and disk component, with the corona component having a slightly larger contribution.
In the jet model, the seed photons are dominated by disk photons.

In the corona model, pair production becomes important at energies around GeV and above (Figure~\ref{fig:sed}).
The ratio of intrinsic to attenuated $\gamma$-ray luminosity at 100~MeV and 1~GeV is about $\ratioI$ and $\ratioII$, respectively.  
Cascade would occur following annihilation of GeV photons with keV photons in the corona, and produce secondary sub-GeV photons due to inverse Compton scattering \citep[e.g.,][]{Murase2012,Murase2020,Murase2024}.
To estimate this effect, we calculate the positron/electron spectra following Eq.~(3.25) in \cite{Aharonian2004} and the subsequent inverse Compton spectra using Eqs.~(\ref{eq:Lnup}-\ref{eq:Lnu}) in the Appendix.
The emergent SED converges after a few iterations, leading to a flux enhancement by a factor of approximately $\ratioIII$ at 100~MeV.
Thus, if the cascade process is considered, a lower electron density by a factor of about $\ratioIII$ would be sufficient to account for the observed data.
An accurate calculation, however, needs to solve the radiative transfer equation by incorporating the photon generation term due to cascade.
Consequently, Eqs.~(\ref{eq:zeta}) and (\ref{eq:Lnu}) are inapplicable in their current form.
This is beyond the scope of this work and will be addressed in the future.

As discussed above, the star formation activity and outflows could be other possible causes of $\gamma$-ray emission.
Star-forming galaxies display significant $\gamma$-ray emission with a luminosity scaled with the infrared luminosity \citep{Ajello2020}, suggesting that their $\gamma$-ray emission is driven by star forming processes, possibly due to interaction between high energy cosmic rays and the interstellar medium \citep{Paglione1996, Abdo2010, Ackermann2012_SF, Ajello2020}.
We estimate the contribution from star formation to $\gamma$-ray emission using the $L_{0.1-800\,\rm GeV}-\LIR$ relation established by \cite{Ajello2020}.
The infrared luminosity, $\LIR=\LIRVal$, is quoted from the IRAS Faint Source Catalogue \citep{Moshir1990}.
The expected $\gamma$-ray luminosity, $\LgamSFG$, is about an order of magnitude below the observed luminosity, indicating that the contribution from star formation activity is negligible.
In theory, $\gamma$-rays can be produced by interaction between AGN-driven outflows and the interstellar medium \citep{Faucher2012,Wang2016,Zhu2024},
AGN torus \citep{Wang2004, Inoue2022},
or clouds \citep{Huang2024}.
Possible observational evidence for $\gamma$-rays produced by ultra fast outflows is reported \citep{Ajello2021,Sakai2025}. 
However, \cite{McDaniel2023} searched galaxies hosting molecular outflow for $\gamma$-ray emission, but found no evidence that the outflow is directly accelerating cosmic rays.
\cite{Dall2023} studied the ionized gas outflows from NGC 3281 using ALMA CO(2--1) observations, and estimated a power of $\dot{E}=(0.045-1.6)\times10^{40}\,\rm erg\,s^{-1}$. This is much lower than the $\gamma$-ray luminosity measured in this work, indicating that the $\gamma$-rays from NGC 3281 can not be dominated by the outflows.
Thus, these processes are unlikely to account for the observed $\gamma$-ray emission seen in NGC 3281.

If the $\gamma$-ray emission is indeed due to a jet, it should appear as a weak version of radio-loud AGN.
A correlation between $\LnuI$ and $L_{\rm 0.1-100\rm GeV}$ has been found for radio-loud AGNs 
\citep{Inoue2011,DiMauro2014,Fukazawa2022}.
Such a relation may extend to radio-quiet AGNs.
We plotted NGC 3281 in the same diagram (Figure~\ref{fig:rgcorr}) and found that it aligns with the correlation, among the lowest luminosity regime in the sample. 
This resembles the Fanaroff–Riley0 (FR0) galaxies, which possess compact radio jets and also show a correlation between their radio and $\gamma$-ray emission consistent with the relation seen in canonical FR radio galaxies \citep{Khatiya2024}.
On the other hand, such a low radio luminosity may have other origins than low-power jets \citep{Panessa2019}.
NGC 3281 has outflows with a power of $(0.045-1.6)\times10^{40}\,\rm erg\,s^{-1}$,
which might contribute to the observed radio flux \citep{Dall2023}.
Synchrotron emission in the corona is another possibility.  
In radio-quiet galaxies, the radio emission follows a trend of $\LnuI/\Lx \sim10^{-5}$, arguably attributed to corona emission \citep{Laor2008,Smith2020}.
NGC 3281 shows a consistent radio-to-X-ray luminosity ratio ($1.3 \times 10^{-5}$, see \S~\ref{sect:jet}), suggesting that one should be cautious about the radio origin. 
In this regard, the jet model is valid only if the radio emission is dominated by relativistic jets. 
In the jet model, we have assumed that the size of the emission region approximates the  distance to the SMBH, parameterized with the same $R$. 
In reality, the emission size ($R_{\rm e}$) may be smaller than the distance ($R_{\rm d}$).
According to Eq.~(\ref{eq:Lnup}), the $\gamma$-ray luminosity has a scaling relation $L_\nu^\prime \propto R_{\rm e}^3 R_{\rm d}^{-2} n_{\rm e}$.
Therefore, one should keep the caveat in mind that if the actual emission size is smaller than the distance, the true electron density will be higher than that quoted in Table~\ref{tab:sed_para}.

\begin{acknowledgments}
We thank Jian Li, Ailing Wang, and Qing-Chang Zhao for useful discussions and the anonymous referee for useful comments.
We acknowledge funding support from the National Natural Science Foundation of China under the grant Nos.\ 12025301, 11991050, 11991054, 12333003, the National Key R\&D Program of China (2021YFA1600404), and the Strategic Priority Research Program of the Chinese Academy of Sciences.
\end{acknowledgments}

\facilities{{\it Fermi}-LAT, Swift-BAT}
\software{{\tt ScienceTools} (v2.0.8) and {\tt FermiPy} \citep{Wood2017}}

\appendix
\restartappendixnumbering
\section{Inverse Compton scattering process}
\label{sect:IC}
In the model of inverse Compton scattering, the $\gamma$-ray spectrum can be expressed as \citep{Blumenthal1970, Inoue1996},
\begin{equation}\label{eq:Lnup}
\Lnup=\Rcom^3h\nu
\int_{\nu_{\rm min}}^{\nu_{\rm max}}
\int_{\gammin}^{\gammax}
f(\nu, \nu_i, \game)
\frac{{\rm d}N}{{\rm d}\game}
n_{\rm ph}(\nu_i)
{\rm d}\game {\rm d}\nu_i,
\end{equation}
where
$\Rcom$ is the size of emission region,
$\nu_i$ is the incident photon frequency,
$\nu_{\rm min}=10^{13}\,\rm Hz$ and $\nu_{\rm max}=10^{21}\,\rm Hz$ are the minimum and maximum of the incident photon frequencies, respectively,
$\nu$ is the $\gamma$-ray frequency,
$\game$ is the electron Lorentz factor,
$dN/d\game$ is the spectrum of the electron in Eq.~(\ref{eq:electron}),
$\gammin$ is the minimum of the electron Lorentz factor,
$n_{\rm ph}(\nu_i)$ is the spectrum of incident seed photons number density derived from the photon field described in \S\,\ref{sect:corona}.
The function
\begin{equation}
f(\nu, \nu_i, \game)=\frac{2\pi r_{\rm e}^2c}{\game^2\nu_i}\left[2\,q\,\ln q +(1+2q)(1-q)+\frac{(\Gamma_e q)^2(1-q)}{2(1+\Gamma_e q)}\right],
\end{equation}
determines the spectrum shape after an incident photon is scattered by a relativistic electron \citep{Blumenthal1970},
where $r_{\rm e}=e^2/\me c^2$ is the classical electron radius,
$e$ is the electron charge,
$\me$ is the electron mass,
$\Gamma_e=4h\nu_i\game/\me c^2$, and $q=\nu/4\game^2\nu_i(1-h\nu/\game \me c^2)$.

\section{pair production process}
\label{append:pair}
\begin{figure}
\centering
\includegraphics[width=0.49\linewidth]{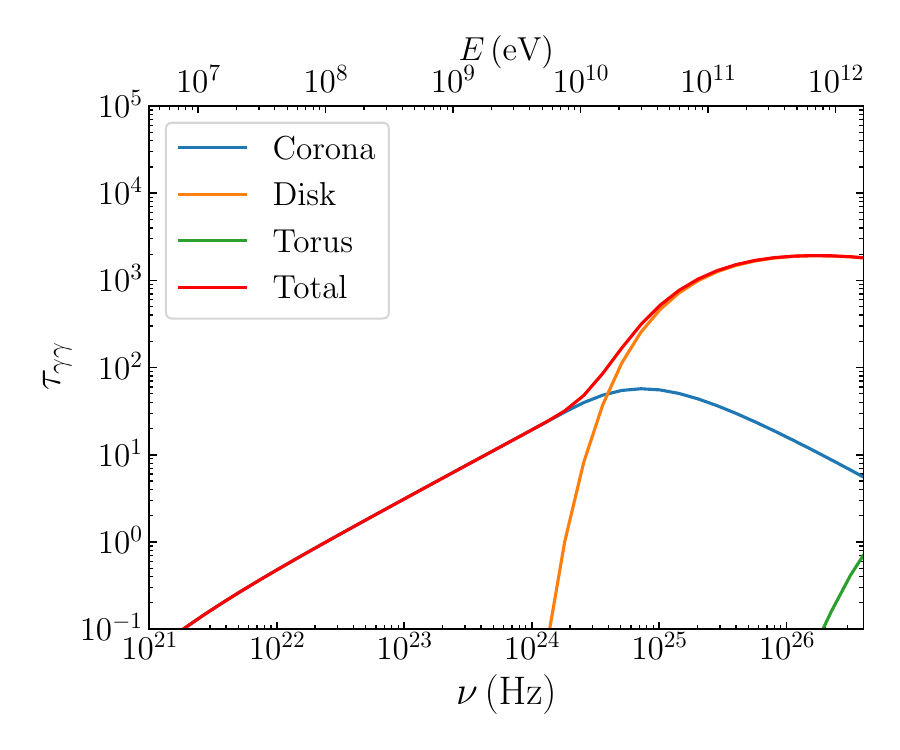}
\includegraphics[width=0.49\linewidth]{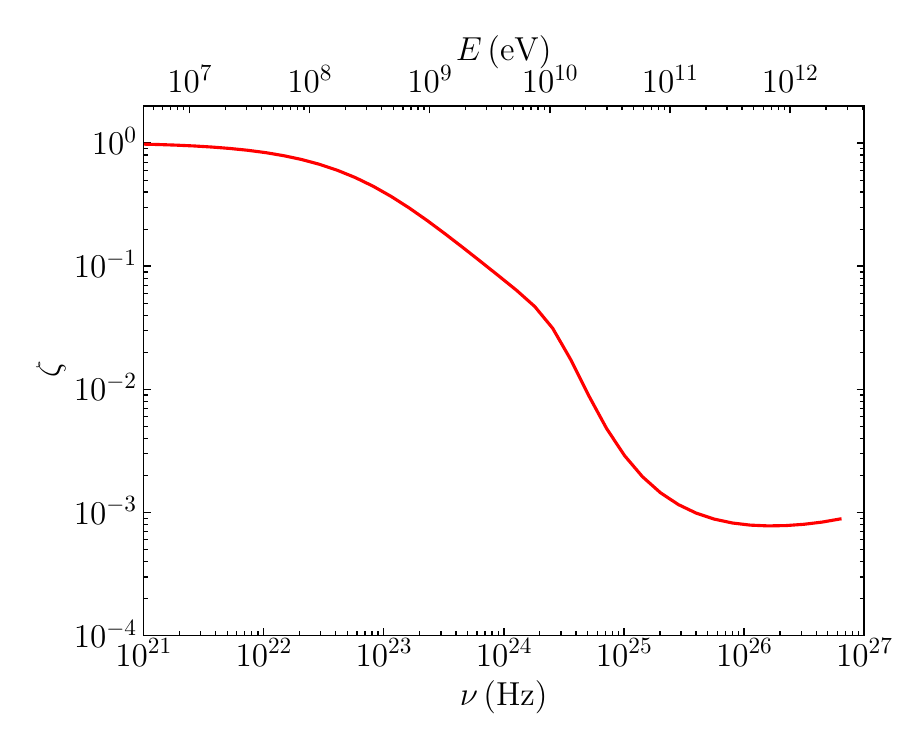}
\caption{Optical depth of pair production (left) and fraction of transmission (right) as a function of $\gamma$-ray frequency/energy for the corona model.}
\label{fig:tau}
\end{figure}

The $\gamma$-rays produced in the corona would be attenuated due to pair production ($\gamma\gamma^\prime\rightarrow e^{+} e^{-}$) interactions.
The total cross section of $\gamma\gamma^\prime$ pair production process is given by \cite{Aharonian2004},
\begin{equation}\label{eq:sigma}
\sigmagg(\nu,\nu_i)=
\frac{3\,\sigmaT}{2 s^{2}}
\left[\left(s+\frac{1}{2}\ln s-\frac{1}{6}+\frac{1}{2\,s}\right)
\ln\left(\sqrt{s}+\sqrt{s-1}\right)-\right.
\left. \left(s+\frac{4}{9}-\frac{1}{9\,s}\right)\sqrt{1-\frac{1}{s}} \right],
\end{equation}
where $\sigmaT$ is the Thomson cross section,
$s\equiv h^2\nu \nu_i/\me^2\,c^4$,
$h\nu$ is the energy of the $\gamma$-rays,
and $h\nu_i$ is the energy of the incident target photon.
$\sigmagg(\nu,\nu_i)$ reaches its maximum value $\approx 0.22 \,\sigmaT$ at $s=3.5$.
For photons with $h\nu=1\rm\, GeV$, the most efficient pair production occurs with target photons with $h\nu_i=0.91$~keV,
which is in the Swift-XRT energy range \citep{Burrows2005}.
The optical depth to pair production is given by
\begin{equation}\label{eq:tau}
\taugg
=\int_{\nu_{\rm min}}^{\nu_{\rm max}}
n_{\rm ph}(\nu_i)\sigmagg(\nu,\nu_i) \Rcom {\rm d}\nu_i
=\int_{\nu_{\rm min}}^{\nu_{\rm max}}
\left(\frac{L_{\nu_i}}{4\pi \Rcom^{2}c h\nu_i}\right)
\sigmagg(\nu,\nu_i) \Rcom {\rm d}\nu_i,
\end{equation}
where
$n_{\rm ph}(\nu_i)$ is the spectrum of the soft photon number density,
and $L_{\nu_i}$ is the soft X-ray luminosity density.
$\taugg$ due to the three components is shown in the left panel of Figure \ref{fig:tau}.
The fraction of high-energy photons remaining after attenuation is given by the factor \citep{Dermer2009},
\begin{equation} \label{eq:zeta}
\zetanu=\left(\frac{3}{\taugg}\right)
\left[
\frac{1}{2}+\frac{{\rm exp}(-\taugg)}{\taugg}-
\frac{1-{\rm exp}(-\taugg)}{\taugg^{2}}
\right] \; ,
\end{equation}
shown in the right panel of Figure \ref{fig:tau}.
Finally, we derive the
result of the attenuated $\gamma$-ray luminosity spectrum
\begin{equation}\label{eq:Lnu}
L_\nu=\zetanu \Lnup.
\end{equation}

\bibliography{corona}{}
\bibliographystyle{aasjournal}

\end{document}